\documentclass[prl,twocolumn,superscriptaddress,nofootinbib]{revtex4-2}
\pdfoutput=1
\usepackage{ulem}
\usepackage{amsmath,bbold}
\usepackage{amssymb}
\usepackage{color}
\usepackage{amsthm}
\usepackage{hyperref}
\usepackage{cleveref}
\usepackage{cancel}
\usepackage{stmaryrd}
\usepackage{soul}
\usepackage{subcaption}
\usepackage{todonotes}
\usepackage{adjustbox}
\definecolor{ma}{RGB}{246,76,246}

\begin{document}

\title{Adversarial Thermodynamics}
\author{Maite Arcos}
\email{maite.arcose@gmail.com}
\affiliation{University College London, London, UK}
\affiliation{Institute of Physics, École Polytechnique Fédérale de Lausanne (EPFL), Lausanne, Switzerland}

\author{Philippe Faist}
\affiliation{Dahlem Center for Complex Quantum Systems, Freie Universität Berlin, Berlin, Germany}

\author{Takahiro Sagawa}
\affiliation{Department of Applied Physics, The University of Tokyo, Tokyo, Japan}
\affiliation{Quantum-Phase Electronics Center (QPEC), The University of Tokyo}

\author{Jonathan Oppenheim}
\affiliation{University College London, London, UK}

\begin{abstract}
Work extraction from finite systems is inherently stochastic, raising the question of how a thermodynamic engine should be operated when both the amount of extracted work and its reliability matter. We address this question by establishing a connection between a Szilard-type engine and Kelly gambling, accounting for the additive nature of thermodynamic work. We derive an explicit family of optimal work-extraction protocols parametrized by an agent's risk preferences, and show that their certainty-equivalent work is given by Rényi divergences, providing an operational interpretation of generalized free energies. Independently, we show that the same family of protocols optimizes the reliability exponents of attaining prescribed work yields, both below and above the maximal average-work rate. Thus, two distinct optimization principles—risk-sensitive expected utility and work-extraction reliability—select the same thermodynamic protocols, connecting generalized free energies to the achievable trade-off between work and reliability.
\end{abstract}

\maketitle

\textit{Introduction.}
The second law of thermodynamics underpins our understanding of heat engines, information processing, and nonequilibrium phenomena. In the thermodynamic limit it can be viewed either as a bound on average entropy production (equivalently, free-energy decrease) or as a constraint on which state transformations are possible.

At small scales, however, work and entropy production fluctuate strongly, and deterministic thermodynamic potentials no longer fully characterize operational performance. This motivates two major modern frameworks: stochastic thermodynamics, which treats work as a fluctuating quantity and constrains entropy production statistically~\cite{Jarzynski1997,seifert2012stochastic,ParrondoHorowitzSagawa2015}, and thermodynamic resource theories, which express the second law as constraints on state transitions and yield families of generalized second laws in terms of Rényi divergences~\cite{janzing_thermodynamic_2000,horodecki_reversible_2003,HO-limitations,ruch1976principle,brandao2015second} (see Ref.~\cite{takahirosbook} for a review). 

While both approaches successfully characterize constraints, they do not directly specify how an agent's tolerance for risk should determine the protocol used to extract work from a finite system. 

Here we address this question by making the work-extraction protocol itself the object of optimization. We identify an explicit protocol-level connection to Kelly gambling~\cite{kellyspaper}, a classic information-theoretic problem of optimizing wealth allocation under uncertainty. In this mapping, Bob's imposed constraint plays the role of the odds, Alice's choice of protocol plays the role of the allocation made before the outcome is known, and the extracted work has the corresponding logarithmic payoff structure. This converts finite-size work extraction into a risk-sensitive protocol-optimization problem. Maximizing expected utility yields an explicit optimal protocol for each risk tolerance.
Notably, the same family of protocols arises from a second, independent route: optimizing the reliability of extracting a prescribed amount of work, without assuming a utility function. Below the maximal average-work rate, this amounts to maximizing the exponent at which the failure probability vanishes; above it, to minimizing the exponent governing the success probability. The two criteria are operationally different: one optimizes expected utility under the true prior, while the other optimizes the reliability of work extraction. Their agreement shows that risk sensitivity and reliability constraints are two ways of specifying the same thermodynamic control principle. This gives the mapping a direct physical interpretation: the agent's risk tolerance, or equivalently a reliability requirement, selects how the engine should be operated.

Concretely, we implement this correspondence in a Szilard-type engine with imposed constraints. The formal analogy to Kelly betting~\cite{kellyspaper} comes with a crucial physical difference: financial wealth compounds multiplicatively, whereas thermodynamic work is additive. As a result, the thermodynamic problem is not obtained by simply translating the gambling problem term by term; the additivity of work selects the appropriate risk-sensitive objective and changes the optimal protocols. Building on recent generalizations of Kelly gambling~\cite{classicalswisspeople,GamblingResourceTheory}, we show that the resulting optimized work values are parametrized by R\'enyi divergences and are achieved by explicit thermodynamic protocols. In particular, the deterministic equivalent assigned to the fluctuating work becomes a generalized free energy, a feature absent from the multiplicative gambling problem.

Several works have connected thermodynamics with gambling and utility-theoretic ideas~\cite{Vinkler2016Analogy,hirono2015jarzynski,ito2016backward,Manzano2021ThermoGamblingDemons,Tohme2024GamblingCarnotEngine,KellyQuantumpPayoff,ducuaraandpaul}. Ref.~\cite{Vinkler2016Analogy} developed a Kelly-type analogy for measurement-based work extraction, where measurements act as side information. Ref.~\cite{KellyQuantumpPayoff} formulated Kelly betting with a quantum payoff, where the gambler's capital is encoded in the ergotropy of a quantum memory. Ref.~\cite{ito2016backward} related gambling and thermodynamics through information flow. Refs.~\cite{Manzano2021ThermoGamblingDemons,Tohme2024GamblingCarnotEngine} considered gambling-inspired protocols in which the gambling element is temporal: the agent waits for a favorable stochastic condition before stopping or quenching the process. Ref.~\cite{ducuaraandpaul} applied expected utility theory to fixed stochastic thermodynamic processes obeying Crooks fluctuation relations. These works use gambling or utility-theoretic ideas to describe side information, information flow, timing of interventions, quantum payoffs, or valuation of a fixed process. To our knowledge, the present work establishes the first protocol-level isomorphism between Kelly gambling and finite-size work extraction.

\textit{The adversarial Szilard engine.}
We begin by formalizing a thermodynamic work extraction problem as a setup between adversaries, mirroring the structure of Kelly betting, which we review in the Supplementary Material. Consider three players: Bob, who sets initial constraints; Alice, who optimizes work extraction (analogous to a gambler allocating bets), and Charlie, a referee who enforces randomness. 

\begin{center}
\begin{figure}[htbp]
    \includegraphics[width=\columnwidth, scale=0.8]{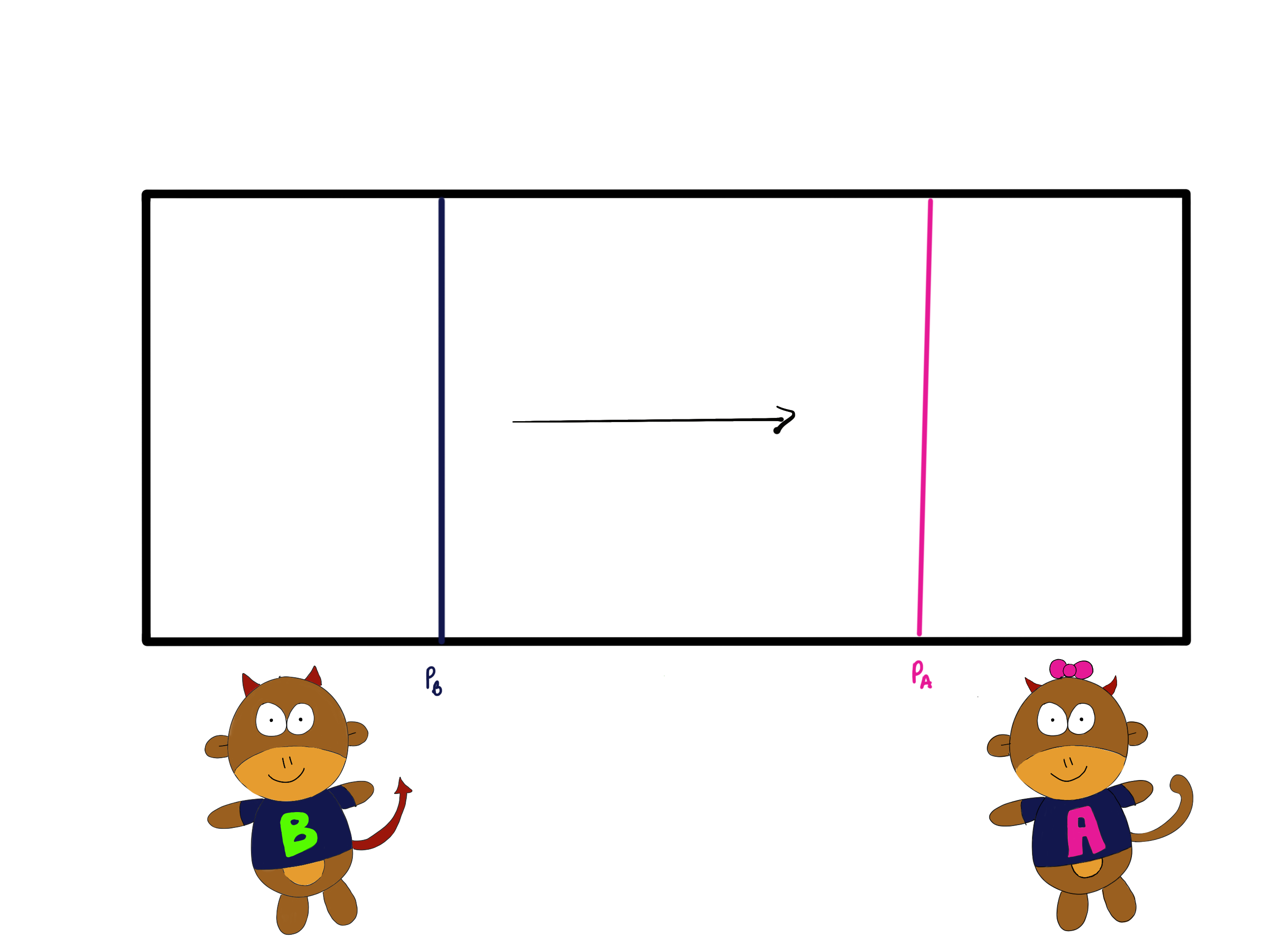}
       \caption{If the molecule is on the left, Alice extracts work while moving the partition to her chosen position $Q^A$. If the molecule is on the right, the protocol requires Alice to do work against the molecule.}
       \label{fig:subC}
\label{AdversarialSzilardEngine}
\end{figure}
\end{center}

We consider an empty box of volume $V$ which Bob divides into two parts by placing a partition at some position $Q^{B}$. The referee, Charlie, then samples from a binary probability distribution $P(x)$ and places a molecule on the left or right-hand side of the box according to the outcome, which Alice and Bob do not know.  
Here, $x=0$ (resp. $1$) represents that the molecule is in the left (resp. right) side.
Alice proceeds to extract work by performing isothermal compressions and expansions on the box, moving the partition to a final position $Q^{A}$ of her choice. The process is illustrated in Fig. \ref{AdversarialSzilardEngine}. 
Since this setup resembles the Szilard engine \cite{szilardengine}, we name it the adversarial Szilard engine.  However, we emphasize that our setup does not involve measurement and feedback by an external ``demon'', in contrast to Szilard's original setup. 
The formulation can be viewed as an abstract model of finite information engines with tunable control parameters. It is close to experimentally realized two-state information engines, such as single-electron-box Maxwell-demon experiments, where the system state is stochastic and the control parameter can be tuned externally~\cite{TakahiroMutualInformationExperiment}. 

Without loss of generality, assume that Bob places the partition closer to the leftmost edge of the box, so that the volume on the left-hand side can be expressed as $Q^{B}V$, and that Alice moves her partition to a position so that the volume on the left-hand side can be expressed as $Q^{A}V$ in terms of fractions $0< Q^{A} < 1 $ and $0< Q^{B}  < 1$ of the total length of the box. 
The work extracted by Alice depends on the molecule's position $x=0,1$:
\begin{equation}
    w (0) = k_{B}T \ln \left( \frac{Q^{A}}{Q^{B}} \right); \ \ 
   w (1) = k_{B}T \ln\left(\frac{1-Q^{A}}{1-Q^{B}}\right).
\end{equation}
To maintain consistency with the notation of the literature \cite{coverandthomas, classicalswisspeople}, we associate each partition position $q$ with the corresponding two-component vector of relative volumes, $q(0):=q$ and $q(1):=1-q$. These relative volumes give the equilibrium distribution for a molecule in a box whose two sides have volumes $qV$ and $(1-q)V$. Thus, when $Q^{A}$ or $Q^{B}$ appears in a relative entropy, it denotes this distribution over the outcomes $x=0,1$, not merely the scalar coordinate of the partition.

The average amount of work $W$ extracted by Alice is given by 
\begin{equation}
W=  k_{B}T\bigg( P(0) \ln \bigg( \frac{Q^{A}(0)}{Q^{B}(0)} \bigg) + P(1)\ln \bigg( \frac{Q^{A}(1)}{Q^{B}(1)} \bigg)  \bigg), 
\end{equation}
or equivalently,
\begin{eqnarray}\label{adversarialengine}
  \boxed{W =k_{B}T \big( D(P||Q^{B}) - D(P||Q^{A}) \big)  }
\end{eqnarray}
where $D ( \cdot \| \cdot )$ is the standard relative entropy (the Kullback-Leibler divergence). 
This formula is true not only for the above binary adversarial Szilard engine (Fig.~\ref{AdversarialSzilardEngine}), but also for general multi-level engines (see SM).

The quantity \eqref{adversarialengine} is maximized when Alice knows the prior $P$ and sets $Q^{A}=P$, in which case $W= k_{B}T  D(P||Q^{B})$. This maximal extractable work from the system can be identified with the nonequilibrium free energy from thermodynamics up to a change in the base of the logarithm. In the particular case where Bob puts the partition in the middle, one obtains $k_{B}T (\ln 2-H(P))$, where $H(\cdot )$ is the Shannon entropy. 
For intuition, it is also useful to consider Charlie placing the molecule uniformly in space. Then $P(0)=Q_B$, so $P=Q_B$ as a Bernoulli distribution over the two sides. Equation~(3) gives $W=-k_BT D(Q_B\|Q_A)\leq0$, with equality only for $Q_A=Q_B$. Thus positive average work is possible only when Charlie's prior is biased relative to the equilibrium distribution induced by Bob's partition.

The expression for the average extracted work by Alice is formally related to a formula appearing in the classical Kelly betting problem from information theory \cite{kellyspaper}: both involve a difference of relative entropies, and are equivalent up to an exponentiation. However, we note that, in Kelly’s gambling problem, wealth evolves multiplicatively, making the natural utility functions those of the CRRA type, as considered in \cite{classicalswisspeople}. 

Equation~\eqref{adversarialengine} also appears in mismatch settings where a device optimized for one distribution is applied to another~\cite{KolchinskyWolpert2017,KolchinskyWolpert2021,KolchinskyWolpert2021b,Wolpert2016FreeEnergyRequirements}. In this Letter we focus on the known-prior setting and use it to connect Eq.~\eqref{adversarialengine} to risk-sensitive decision theory; extensions to unknown priors can be treated within the mismatch framework. When Alice does not know the correct prior, our adversarial Szilard engine reduces to the formulation of  \cite{KolchinskyWolpert2017} at the level of the ensemble average.

Importantly, the primary focus of the subsequent arguments will be connecting Eq. \eqref{adversarialengine} to expected utility theory, economics, and Kelly betting, all of which assume a known distribution in the ideal case. 
Therefore, we will assume that the prior distribution is known to Alice. A brief review of Kelly betting is given in Supplementary Material. 

\textit{Utility maximization: connection to expected utility formulations.}
So far, we showed that the adversarial Szilard engine leads to an optimization problem closely related to Kelly betting, matching its structure up to an exponentiation. We next use this observation to connect work extraction to expected utility theory and to the generalized free energies that appear in resource-theoretic thermodynamics.

Risk sensitivity is encoded in expected utility theory through the choice of utility function, and rational decisions maximize expected utility rather than expected value~\cite{vonneumann1944gametheory}. In Kelly betting, where wealth compounds multiplicatively, it is natural to parametrize risk preferences using CRRA utilities~\cite{classicalswisspeople}. Thermodynamic work, however, is additive; correspondingly, the relevant class is CARA utilities, which have also appeared in risk-sensitive thermodynamic valuations and related fluctuation-theorem settings~\cite{ducuaraandpaul,FrancicaDellAnna2024}. We therefore model an agent’s protocol choice by expected-utility maximization, which selects the optimal work-extraction strategy as a function of a single risk parameter. 

The utility function we are considering (CARA) is given by 
\begin{equation}\label{CARAutility}
    u_{r}(w)= \frac{ 1- e^{-rw/k_BT}}{r},
\end{equation}
where $w$ is the work as a random variable. 
The parameter $r$ reflects the risk preferences of the agent: 
\begin{itemize}
\item 
$r>0$: \textit{Risk-aversion} (concave utility) --- The agent prefers guaranteed work extraction over uncertain fluctuations. Here risk aversion increases with increasing $r$. 
\item 
$r=0$: \textit{Risk-neutral} behavior (linear utility, recoverable via limit $r \rightarrow 0$) --- The agent maximizes expected work regardless of fluctuations.
\item $r<0$: \textit{Risk-seeking} behavior (convex utility) --- The agent prefers higher amount of work extracted even if it means lower probability of success.
\end{itemize}
 We now show that maximizing expected utility over protocols yields explicit optimal strategies, Eq.~\eqref{optimalstrategy}, and that both the certainty equivalent and expected work are given by Rényi divergences, Eqs.~\eqref{eq:WorkCE} and~\eqref{eq:expected}.         
For a single round of work extraction in the adversarial Szilard engine described above, the extracted work in terms of the outcome $x$ can be expressed as
\begin{equation}\label{generalsinglework}
    w(x) = k_{B}T \ln \bigg( \frac{Q^{A}(x)}{Q^{B}(x)} \bigg) .
\end{equation} 
It follows using Eq.~\eqref{CARAutility} that the utility of outcome $x$ is 
\begin{equation}\label{generalsinglework_u}
    u_{r}(w(x))= \frac{1}{r} \bigg(1-  \bigg(  \frac{Q^{A}(x)}{Q^{B}(x)}\bigg)^{-r} \bigg).
\end{equation}
See Supplementary Material  for the case of multi-level engines.
By the expected utility hypothesis, a rational agent will set their bet $Q^{A}$ as to maximize 
\begin{equation}\label{generalsinglework_u2}
    \sum_{x} P(x) u_{r}(w(x)) = \frac{1}{r} \bigg(1- \sum_{x}P(x) \bigg(  \frac{Q^{A}(x)}{Q^{B}(x)}\bigg)^{-r} \bigg).
\end{equation}
We show that for a given $r$,  the optimal choice of $Q^{A}$ corresponds to 
\begin{equation}\label{optimalstrategy}
    Q^{A,r}= \frac{P(x)^{\frac{1}{1+r}} Q^{B}(x)^{\frac{r}{1+r}}}{ \sum_{x'}P(x')^{\frac{1}{1+r}} Q^{B}(x')^{\frac{r}{1+r}}} ,
\end{equation}
where we have introduced the notation $Q^{A,r}$ to indicate that the optimizer depends on the risk aversion of the agent (see SM for details). 
The expression above gives the interior optimum for $r>-1$. In the risk-neutral limit $r\to 0$, $Q^{A,r}$ goes to $P$, while in the limit $r\to\infty$ of extreme risk aversion, $Q^{A,r}$ goes to $Q^B$, corresponding to not moving the partition and hence avoiding fluctuations. For sufficiently strong risk seeking, the optimum is instead attained at the boundary. In particular, in the limit $r\to -\infty$ and for $Q^B<1/2$, Alice chooses $Q^A\to 1$, maximizing the favorable work outcome when the molecule is on the left, at the cost of an arbitrarily large work investment if the molecule is on the right.

In our adversarial Szilard engine, the strategy $Q^{A,r}$ corresponds directly to Alice’s choice of partition placement; in economic terms, it reflects the unique optimal allocation for a CARA agent with parameter $r$. 
In the following, we connect $Q^{A,r}$ to the average work extracted by an agent with a particular level of risk aversion, as well as to the Rényi divergences (the generalized free energies of the resource-theoretic approach).

The corresponding average extracted work for risk aversion level $r$ is given by 
\begin{equation}
    \boxed{ W^{(r)}=  k_{B}T \left( \frac{1}{1+r} D(P||Q^{B}) + \frac{r}{1+r}D_{\frac{1}{1+r}}(P||Q^{B}) \right) }
    \label{eq:expected}
\end{equation}
(see SM for full derivation).
Here, we defined the Rényi divergence for $-\infty < \alpha < \infty$ following Ref.~\cite{vanerven2014} :
\begin{equation} \label{def_Rényi}
D_\alpha(P || Q^B) = \frac{1}{\alpha - 1} \ln\sum_x P(x)^\alpha Q^B(x)^{1-\alpha}.
\end{equation}
 Eq.~\eqref{eq:expected} clarifies that the ensemble average of the extracted work depends on the particular strategy and level of risk aversion. 
Noticing that $\frac{1}{1+r} + \frac{r}{1+r} = 1$, Eq.~\eqref{eq:expected} is a weighted sum of $D(P||Q^{B})$ and $D_{\frac{1}{1+r}}(P||Q^{B})$.

An equivalent way to summarize risk sensitivity is via the certainty equivalent $W_{CE}$, defined as the deterministic work value that makes the agent indifferent to the fluctuating outcome, i.e., the value such that $u_{r}(W_{CE})= \mathbb E(u_{r}(w(x)))$, where $\mathbb E$ denotes the ensemble average. We show (see SM) that the certainty equivalent of extracted work for an individual with risk aversion level $r$ is given by the Rényi divergence
\begin{equation}
    \boxed{W_{CE}= k_{B}T D_{\frac{1}{1+r}} (P||Q^{B}) } 
    \label{eq:WorkCE}
\end{equation}
We note that the certainty equivalent for $n$ rounds is exactly $n$ times the single-round certainty equivalent due to the additivity of work and independence of rounds.

We also note that an analogous expression for the certainty equivalent does not hold in the context of gambling, because the utility function in that case is different. 
A related but distinct approach was taken in Ref.~\cite{ducuaraandpaul}, which considered a fixed stochastic thermodynamic process satisfying the Crooks fluctuation theorem and quantified the certainty equivalent of the dissipated fluctuating work. They found
\begin{align}
W_{CE}^{\mathrm{diss},r}
= k_{B}T  D_{1+r}\big(P_{\mathrm{F}}(w),||,P_{\mathrm{R}}(-w)\big),
\label{eq:WorkCWEaul}
\end{align}
where $P_{\mathrm F}(w)$ and $P_{\mathrm R}(-w)$ are the forward and reverse work distributions, respectively.
The comparison with our result should not be overinterpreted. In Ref.~\cite{ducuaraandpaul}, the Rényi divergence compares forward and reverse work distributions associated with a fixed Crooks process. In contrast, in Eq.~\eqref{eq:WorkCE} the distributions are Charlie's prior $P$ and the left/right distributions associated with the partition positions $Q^A,Q^B$. Moreover, our adversarial Szilard engine is not formulated as a Crooks forward--reverse pair: the central task is to optimize the control variable $Q^A$, thereby choosing the work distribution itself. Thus, while both approaches show that Rényi divergences arise naturally when incorporating risk sensitivity into thermodynamics, they provide distinct operational interpretations.

It is useful to compare the expected extracted work and the certainty equivalent (CE): the former captures the stochastic-thermodynamic viewpoint of average work, while the latter provides an operational, risk-sensitive guarantee that connects naturally to generalized free energies. Their joint behavior therefore clarifies how risk attitudes interpolate between these viewpoints and provides a basic rationality check on the agent’s strategy. In our adversarial Szilard engine, placing the partition proportional to probability corresponds to risk neutrality ($r=0$), for which $\mathrm{CE}=\mathbb{E}[W]$ and both reduce to the nonequilibrium free energy.

For risk aversion ($r>0$), the CE is $D_{\frac{1}{1+r}}(P\|Q^B)$ [Eq.~\eqref{eq:WorkCE}] and decreases monotonically with $r$ by R\'enyi monotonicity, reflecting the price paid for certainty. For risk seeking ($r<0$), strategies may have negative expected work (net dissipation) despite rare positive outcomes; however, the CE remains consistent with first-order stochastic dominance, i.e., it never prescribes a sure outcome that is strictly worse than every possible realization of the gamble (see Supplementary Material for details).

\textit{Thermodynamic risk aversion.}
We have shown that the nonequilibrium free energy from thermodynamics can be understood as the maximum amount of work that can be extracted in the adversarial engine at the level of the ensemble average. We next analyze the same ensemble from a finite-size regime perspective using tools developed in \cite{csiszar1998method, GamblingResourceTheory}, and to prove that work extraction in the finite-size regime corresponds to the maximization of a utility function as described before.

Recall first that for a single round of work extraction, the extracted work in terms of the outcome $x$ of the random variable is expressed by Eq.~\eqref{generalsinglework}.
It follows that for $n$ rounds of i.i.d., the work extracted is given by
\begin{equation}
      W_{n} (x^n)= k_{B}T  \sum_{i=1}^{n} \ln \bigg( \frac{Q^{A}(x_i)}{Q^{B}(x_i)} \bigg),
\end{equation}
where $x_i$ is the outcome of the $i$th round and $x^n = (x_1, x_2, \cdots, x_n)$ describes the realized sequence.
By grouping terms with the same outcome, and denoting by $N_{x^n}(x)$ the number of times each outcome occurs in $x^n$, this can be expressed as 
\begin{equation}
      W_{n}(x^n)= k_{B}T n\sum_{x} \lambda_{x^n}(x) \ln \bigg( \frac{Q^{A}(x)}{Q^{B}(x)} \bigg)  .
\end{equation}
It is useful to introduce the type $\lambda_{x^n}$ of the sequence $x^n$~\cite{csiszar1998method}. Often called its empirical distribution, the type is defined as $\lambda_{x^n}(x) = N_{x^n}(x)/n$.  Rewriting the equation above in terms of the type, we obtain an equation analogous to Eq.~\eqref{adversarialengine} in the finite-size regime:
\begin{equation} \label{finite_n_main}
  \boxed{ W_{n} (x^n)=  k_{B}T n(D(\lambda_{x^{n}}||Q^{B})-D(\lambda_{x^{n}}||Q^{A})) }
\end{equation}

Hence, by placing the partition at position $Q^{A}=\lambda_{x^{n}}$ (when Alice's allocation matches the type of the observed sequence), the work extracted is given by $ k_{B}T nD(\lambda_{x^{n}}||Q^{B})$. More generally, if Alice chooses a partition $Q^A$ that does not coincide with the realized type, she still extracts the work given by Eq.~\eqref{finite_n_main}, which is reduced by the mismatch term $D(\lambda_{x^n}||Q^A)$. In the finite-size regime, the work extraction problem can therefore be viewed as a decision-theoretic problem of targeting an empirical type, or a corresponding work/reliability criterion. Each possible type corresponds to a particular fluctuation with probability decaying exponentially up to sub-exponential factors,
while the potential work reward scales as $n D(\lambda_{x^n} || Q^B)$.
This creates a fundamental risk--reward trade-off: Alice must select her protocol to balance the probability of successfully extracting a prescribed amount of work against the magnitude of that work.

To formalize this trade-off, consider a target extracted work \(nk_BT R\), and let \(K=D(P\|Q^B)\) denote the optimal average-work rate. For \(0<R<K\), the relevant optimization is to maximize the exponent at which the probability of failing to attain the target vanishes. For targets above \(K\), within the range specified in the Supplementary Material, the corresponding optimization minimizes the exponent governing the probability of successfully attaining the target.

As we show formally in the Supplementary Material, both optimizations select the family of protocols
\[
Q^{A*,\mu}(x)=\frac{P(x)^\mu Q^B(x)^{1-\mu}}{\sum_{x'}P(x')^\mu Q^B(x')^{1-\mu}},
\]
where \(0<\mu<1\) corresponds to the below-\(K\) regime and \(\mu>1\) to the above-\(K\) regime. In each case, the target rate is \(R=D(Q^{A*,\mu}\|Q^B)\), and the optimal probability exponent is \(D(Q^{A*,\mu}\|P)\).

Notably, this is precisely the family of protocols obtained independently from expected-utility maximization, Eq. \(\eqref{optimalstrategy}\), under the identification \(\mu=1/(1+r)\). Thus, reliability-constrained optimization and risk-sensitive expected-utility maximization select the same thermodynamic protocols.

Together, the expected-utility and reliability formulations give an operational interpretation of the Rényi divergences appearing in this finite engine. In the expected-utility formulation, the optimized protocols realize certainty-equivalent work values given by generalized free energies, Eq. \(\eqref{eq:WorkCE}\). Independently, the same protocols optimize the reliability--yield trade-off at exponential order, as established in the Supplementary Material. Thus, the Rényi divergences characterize not only constraints on thermodynamic transformations, but also the achievable performance of explicit protocols selected according to an agent's risk preferences or reliability requirements.

Our approach reveals that all Rényi divergences $D_\alpha$ acquire an operational interpretation in terms of work extraction for different risk tolerances. Previous works \cite{inadequacyofvnentropy} provided operational interpretation for min and max entropies as bounds in finite-size work extraction. Here, we find a complementary interpretation for all Rényi divergences. Crucially, we also provide the work extraction strategy given the agent's constraints. 

\textit{Conclusions.}
We formulated finite-size work extraction as a risk-sensitive decision problem in which an agent chooses a protocol under constraints and fluctuating outcomes. Because thermodynamic work is additive, this framework naturally selects CARA utilities (in contrast to the CRRA utilities associated with multiplicative Kelly betting), and yields explicit optimal protocols as a function of a single risk parameter. Within this decision-theoretic formulation, both the certainty equivalent (guaranteed) extracted work and the corresponding risk-sensitive expected work are expressed in terms of generalized free energies. 
In the resource-theoretic second laws, these free energies appear as bounds: necessary conditions on which state transformations are possible. Our results supply the achievability side. Each generalized free energy is the guaranteed (certainty-equivalent) work of an explicit, risk-indexed protocol, so the bound is not merely saturated but accompanied by the operation that saturates it, with the risk parameter setting how far along the family one operates. Having explicit optimal protocols for each risk tolerance opens the door to designing nanoscale engines with prescribed reliability. 
This perspective is relevant to experimentally accessible information engines, such as single-electron-box realizations, where the state is stochastic and external control parameters can be tuned.

\textit{Acknowledgments.}
MA thanks Ian Ford and Paul Skrzypczyk for useful discussions. MA was supported by the Mexican
National Council of Science and Technology (CONACYT), EPSRC, the Simons Foundation It from Qubit Network, and SNF Quantum Flagship Replacement Scheme (Grant No. 215933). PhF is supported by the project FOR
2724 of the Deutsche Forschungsgemeinschaft (DFG). TS is supported by JST ERATO Grant Number JPMJER2302, Japan. JO was supported by EPSRC.

\appendix

\bibliography{bibliography,refthermo}

@article{classicalswisspeople,
	Author = {Cédric Bleuler and Amos Lapidoth and Cristoph Pfister },
	Journal = {Entropy},
	Title = { Conditional Renyi Divergences and Horse Betting },
    volume = {22},
    pages = {316},
	Year = {2020}}

@article{kellyspaper,
	Author = { J. Kelly },
	Journal = {Bell System Technical Journal},
	Title = {A new interpretation of the information rate  },
	Year = {1956}
	}

@article{inadequacyofvnentropy,
	Author = {Oscar C. O. Dahlsten and Renato Renner and Elisabeth Rieper and Vlatko Vedral},
	Journal = { New Journal of Physics},
	Title = {Inadequacy of Von Neumann entropy for characterising extractable work },
	 volume    = {13},
  number    = {},
  pages     = {053015},
  year      = {2011}
	}

@book{coverandthomas,
  title={Elements of information theory},
  author={Thomas Cover and Joy Thomas},
  year={2006},
  publisher={Wiley and Sons}
}

@article{szilardengine,
	Author = { L. Szilard },
	Journal = {Z. Phys.},
	Title = { über die Entropieverminderung in einem thermodynamischen System bei Eingriffen intelligenter Wesen },
    volume = {53},
    pages = {840},
	Year = {1929}
	}

@article{ducuaraandpaul,
  title     = {Maxwell's Demon Walks into Wall Street: Stochastic Thermodynamics Meets Expected Utility Theory},
  author    = {Ducuara, Andres F. and Skrzypczyk, Paul and Buscemi, Francesco and Sidajaya, Peter and Scarani, Valerio},
  journal   = {Phys. Rev. Lett.},
  volume    = {131},
  number    = {19},
  pages     = {197103},
  year      = {2023},
  month     = nov,
  publisher = {American Physical Society},
  doi       = {10.1103/PhysRevLett.131.197103},
  url       = {https://link.aps.org/doi/10.1103/PhysRevLett.131.197103}
}

@article{Manzano2021ThermoGamblingDemons,
  title        = {Thermodynamics of Gambling Demons},
  author       = {Gonzalo Manzano and Diego Subero and Olivier Maillet and Rosario Fazio and Jukka P. Pekola and Édgar Roldán},
  journal      = {Phys. Rev. Lett.},
  volume       = {124},
  number       = {23},
  pages        = {230603},
  year         = {2020},
  publisher    = {American Physical Society},
  doi          = {10.1103/PhysRevLett.124.230603}
}

@article{Tohme2024GamblingCarnotEngine,
  title        = {Gambling Carnot Engine},
  author       = {Tarek Tohme and Valentina Bedoya and Costantino di Bello and Léa Bresque and Gonzalo Manzano and Édgar Roldán},
  journal      = {Phys. Rev. Lett.},
  volume       = {131},
  number       = {057101},
  year         = {2023},
  publisher    = {American Physical Society},
  doi          = {10.1103/PhysRevLett.131.057101}
}

@article{KellyQuantumpPayoff,
  title   = {Kelly betting with quantum payoff: a continuous variable approach},
  author  = {Salvatore Tirone and Maddalena Ghio and Giulia Livieri and Vittorio Giovannetti and Stefano Marmi},
  journal = {Quantum},
  volume  = {5},
  pages   = {545},
  year    = {2021},
  doi     = {10.22331/q-2021-09-21-545},
  eprint  = {2001.11395},
  archivePrefix = {arXiv},
  primaryClass  = {quant-ph},
  url     = {https://doi.org/10.22331/q-2021-09-21-545}
}

@article{vanerven2014,
  author    = {Tim van Erven and Peter Harremo{\"e}s},
  title     = {R{\'{e}}nyi Divergence and Kullback-Leibler Divergence},
  journal   = {IEEE Transactions on Information Theory},
  volume    = {60},
  number    =7,
  pages     = {3797--3820},
  year      = {2014},
  month     = jul,
  doi       = {10.1109/TIT.2014.2320500},
  url       = {https://doi.org/10.1109/TIT.2014.2320500}
}

@article{FrancicaDellAnna2024,
  title        = {Optimal work extraction from quantum batteries based on the expected utility hypothesis},
  author       = {Gianluca Francica and Luca Dell’Anna},
  journal      = {Physical Review E},
  volume       = {109},
  number       = {4},
  pages        = {044119},
  year         = {2024},
  month        = apr,
  doi          = {10.1103/PhysRevE.109.044119},
  note         = {Received 27 November 2023; revised 5 February 2024; accepted 11 March 2024; published 8 April 2024}
}

@article{KolchinskyWolpert2017,
  author    = {Kolchinsky, Artemy and Wolpert, David H.},
  title     = {Dependence of dissipation on the initial distribution over states},
  journal   = {Journal of Statistical Mechanics: Theory and Experiment},
  year      = {2017},
  volume    = {2017},
  number    = {8},
  pages     = {083202},
  doi       = {10.1088/1742-5468/aa7e6a},
  url       = {https://doi.org/10.1088/1742-5468/aa7e6a}
}

@article{KolchinskyWolpert2021,
  author    = {Kolchinsky, Artemy and Wolpert, David H.},
  title     = {Work, Entropy Production, and Thermodynamics of Information under Protocol Constraints},
  journal   = {Physical Review X},
  year      = {2021},
  volume    = {11},
  number    = {},
  pages     = {041024},
  doi       = {https://doi.org/10.1103/PhysRevX.11.041024},
  url       = {https://doi.org/10.1103/PhysRevX.11.041024}
}

@article{KolchinskyWolpert2021b,
  author    = {Kolchinsky, Artemy and Wolpert, David H.},
  title     = {Entropy production given constraints on the energy functions},
  journal   = {Physical Review E},
  year      = {2021},
  volume    = {104},
  number    = {},
  pages     = {034129},
  doi       = {https://doi.org/10.1103/PhysRevE.104.034129},
  url       = {https://doi.org/10.1103/PhysRevE.104.034129}
}

@article{GamblingResourceTheory,
  title   = {A Resource Theory of Gambling},
  author  = {Maite Arcos and Renato Renner and Jonathan Oppenheim},
  journal = {arXiv preprint arXiv:2510.08418},
  year    = {2025},
  eprint  = {2510.08418},
  archivePrefix = {arXiv},
  primaryClass = {quant-ph},
  doi     = {10.48550/arXiv.2510.08418},
  url     = {https://arxiv.org/abs/2510.08418},
}

@article{hirono2015jarzynski,
  title={Jarzynski-type equalities in gambling: role of information in capital growth},
  author={Hirono, Yuji and Hidaka, Yoshimasa},
  journal={Journal of Statistical Physics},
  volume={161},
  number={3},
  pages={721--742},
  year={2015},
  publisher={Springer}
}

@article{ito2016backward,
  author  = {Ito, Sosuke},
  title   = {Backward transfer entropy: Informational measure for detecting hidden Markov models and its interpretations in thermodynamics, gambling and causality},
  journal = {Scientific Reports},
  volume  = {6},
  pages   = {36831},
  year    = {2016},
  doi     = {10.1038/srep36831}
}

@article{seifert2012stochastic,
  title={Stochastic thermodynamics, fluctuation theorems and molecular machines},
  author={Seifert, Udo},
  journal={Reports on Progress in Physics},
  volume={75},
  number={12},
  pages={126001},
  year={2012},
  publisher={IOP Publishing}
}

@article{ParrondoHorowitzSagawa2015,
  title={Thermodynamics of information},
  author={Juan M.R. Parrondo and Jordan M. Horowitz and Takahiro Sagawa},
  journal={Nature Physics},
  volume={11},
  number={2},
  pages={131-139},
  year={2015},
  publisher={Nature Publishing Group UK London}
}

@article{brandao2015second,
  title={The second laws of quantum thermodynamics},
  author={Brandão, Fernando and Horodecki, Micha{\l} and Ng, Nelly and Oppenheim, Jonathan and Wehner, Stephanie},
  journal={Proceedings of the National Academy of Sciences},
  volume={112},
  number={11},
  pages={3275--3279},
  year={2015},
  publisher={National Academy of Sciences}
}

@article{csiszar1998method,
  author    = {Imre Csisz{\'a}r},
  title     = {The Method of Types},
  journal   = {IEEE Transactions on Information Theory},
  volume    = {44},
  number    = {6},
  pages     = {2505--2523},
  year      = {1998},
  doi       = {10.1109/18.720546},
  url       = {https://ieeexplore.ieee.org/document/720546}
}

@book{vonneumann1944gametheory,
  title={Theory of Games and Economic Behavior},
  author={Von Neumann, John and Morgenstern, Oskar},
  year={1944},
  publisher={Princeton University Press},
  address={Princeton, NJ},
  edition={1st}
}

@article{Wolpert2016FreeEnergyRequirements,
  title   = {The Free Energy Requirements of Biological Organisms; Implications for Evolution},
  author  = {Wolpert, David H.},
  journal = {Entropy},
  year    = {2016},
  volume  = {18},
  number  = {4},
  pages   = {138},
  doi     = {10.3390/e18040138}
}

@article{Jarzynski1997,
  author    = {Jarzynski, C.},
  title     = {Nonequilibrium Equality for Free Energy Differences},
  journal   = {Physical Review Letters},
  volume    = {78},
  number    = {14},
  pages     = {2690--2693},
  year      = {1997},
  month     = apr,
  doi       = {10.1103/PhysRevLett.78.2690}
}

@article{Vinkler2016Analogy,
author    = {Vinkler, Dror A. and Permuter, Haim H. and Merhav, Neri},
title     = {Analogy between gambling and measurement-based work extraction},
journal   = {Journal of Statistical Mechanics: Theory and Experiment},
volume    = {2016},
number    = {04},
pages     = {043403},
year      = {2016},
month     = apr,
doi       = {10.1088/1742-5468/2016/04/043403},
publisher = {IOP Publishing}
}

@article{TakahiroMutualInformationExperiment,
  author        = {Koski, Jonne V. and Maisi, Ville F. and Sagawa, Takahiro and Pekola, Jukka P.},
  title         = {Experimental study of mutual information in a Maxwell Demon},
  journal       = {Physical Review Letters},
  volume        = {113},
  pages         = {030601},
  year          = {2014},
  doi           = {10.1103/PhysRevLett.113.030601},
  archivePrefix = {arXiv},
  eprint        = {1405.1272},
  primaryClass  = {cond-mat.stat-mech}
}

@book{MaasColell,
  author    = {Mas-Colell, Andreu and Whinston, Michael D. and Green, Jerry R.},
  title     = {Microeconomic Theory},
  publisher = {Oxford University Press},
  address   = {New York},
  year      = {1995},
  isbn      = {9780195102680}
}

@misc{takahirosbook,
  title={Entropy, Divergence, and Majorization in Classical and Quantum Thermodynamics},
  author={Sagawa, Takahiro},
  series={Springer Briefs in Mathematical Physics},
  volume={16},
  year={2022},
  publisher={Springer},
}

@STRING{IEEE = {IEEE}}

@article{ruch1976principle,
	Author = {Ruch, Ernst and Mead, Alden},
	Journal = {Theoretical Chemistry Accounts: Theory, Computation, and Modeling (Theoretica Chimica Acta)},
	Number = {2},
	Pages = {95--117},
	Publisher = {Springer},
	Title = {The principle of increasing mixing character and some of its consequences},
	Volume = {41},
	Year = {1976}}

@article{HO-limitations,
	Author = {Micha\l Horodecki and Oppenheim, Jonathan},
	Eprint = {arXiv:1111.3834},
	Journal = {Nature Communications},
	Number = {2059},
	Title = {Fundamental limitations for quantum and nanoscale thermodynamics},
	Volume = {4},
	Year = 2013}

@article{horodecki_reversible_2003,
	Author = {Horodecki, Micha{\l} and Horodecki, Pawe{\l} and Oppenheim, Jonathan},
	Journal = {Phys. Rev. A},
	Month = jun,
	Number = {6},
	Pages = {062104},
	Title = {Reversible transformations from pure to mixed states and the unique measure of information},
	Volume = {67},
	Year = {2003}}

@article{janzing_thermodynamic_2000,
	Author = {Janzing, D. and Wocjan, P. and Zeier, R. and Geiss, R. and Beth, Th.},
	Journal = {Int. J. Theor. Phys.},
	Month = dec,
	Number = {12},
	Pages = {2717--2753},
	Shorttitle = {Thermodynamic Cost of Reliability and Low Temperatures},
	Title = {Thermodynamic Cost of Reliability and Low Temperatures: Tightening Landauer's Principle and the Second Law},
	Volume = {39},
	Year = {2000}}

\appendix
\clearpage





\section{Review of Kelly betting}\label{kellyreview}

The Kelly betting scheme is a betting setup in which the gambler, Alice, allocates fractions of her wealth to the possible outcomes of a horse race, always distributing a non-zero fraction to each outcome (in order to avoid ever being completely broke). The adversary Bob also sets odds for each possible outcome. 
 
 Suppose that the horse race is described by a random variable $\{ x \}$. Bob, sets the odds $o_{x}$ to outcome $x$, and Alice allocates the fraction $f_{x}$ of her wealth to outcome $x$. With this convention for describing odds, after a round of the game, Alice's initial wealth $\mathcal W_{i}$ is multiplied by the factor $f_{x}o_{x}$: 
 \begin{equation}
     \mathcal W_{1}= f_{x}o_{x} \mathcal W_{i}.
 \end{equation}
 Assuming that Alice reinvests whatever wealth she has after the previous round, her wealth evolves recursively, and at round $N$ it obeys the equation 
\begin{equation}
   \mathcal  W_{n}=  f_{x}o_{x}\mathcal  W_{n-1}.
\end{equation}
Since Alice reinvests her wealth after each round, her wealth after $n$ rounds can be expressed as a product of the wealth multipliers for each outcome:
\begin{equation}
    \frac{\mathcal W_{n}}{\mathcal W_{i}}= \prod_{x} (f_{x}o_{x})^{N_{x^n}(x)},
\end{equation}
 where $N_{x^n}(x)$ is the number of times that the outcome $x$ occurred in the realized sequence $x^n = (x_1, x_2, \cdots, x_n)$. 

Since the fraction by which the wealth is multiplied after every round is a random variable, it is customary to write $Q^{B}(x)=o_{x}^{-1}$ and $Q^{A}(x)=f_{x}$. 
The superscripts $A$ and $B$ indicate that the fraction is chosen by Alice and Bob, respectively. Even though the bets and odds are the same at every round of gambling, the amount of money made by Alice is a random variable. The expression above now becomes

\begin{equation}
    \mathcal W_{n}= \prod_{x} \bigg( \frac{Q^{A}(x)}{Q^{B}(x)} \bigg) ^{N_{x^n}(x)}\mathcal W_{i}.
\end{equation}
 In the limit as $n \gg 1$, the ratio of Alice's initial wealth $W_{i}$ to her final wealth $\mathcal W_{F}$ satisfies
\begin{equation}\label{KellyCT}
    \frac{\mathcal W_{F}}{\mathcal W_{i}} = \exp(n (D(P||Q^{B}) - D(P||Q^{A})) ).
\end{equation}
 Since the relative entropy is non-negative, Alice's optimal strategy is to set \( Q^A = P \), ensuring that her wealth grows at the maximum possible rate. The expression above is the well-known result by Kelly \cite{kellyspaper}. 

As emphasised in the main text, this multiplicative growth of wealth under Kelly gambling contrasts with the additive accumulation of work in the thermodynamic engine, fundamentally shaping the respective utility functions (CRRA vs. CARA) that describe rational behavior in each domain.

\section{Expected utility theory}\label{EUT}

Here we review the concepts from expected utility theory used in the main text. More detailed introductions can be found, for example, in Ref.~\cite{MaasColell}.

Expected utility theory describes decision making under uncertainty. A lottery is a random payoff: it gives payoff $x$ with probability $p_x$.  The expected payoff of the lottery is
\begin{equation}
\langle x\rangle=\sum_x p_x x .
\end{equation}
Expected utility theory assumes that an agent does not necessarily rank lotteries by their expected payoff, but instead by their expected utility,
\begin{equation}
\langle u(x)\rangle=\sum_x p_x u(x),
\end{equation}
where $u(x)$ is the agent's utility function. The utility function encodes the agent's attitude towards fluctuations. A risk-neutral agent has a linear utility function and therefore maximizes expected payoff. A risk-averse agent has a concave utility function, so that
\begin{equation}
\sum_x p_x u(x) \leq u!\left(\sum_x p_x x\right),
\end{equation}
and therefore prefers, all else being equal, a certain payoff to a fluctuating payoff with the same mean. A risk-seeking agent has a convex utility function and prefers fluctuations in the opposite sense.

A useful way to summarize the value of a lottery is its certainty equivalent. For a utility function, the certainty equivalent of a lottery is the deterministic payoff that the agent regards as equivalent to the lottery. It is defined by
\begin{equation}
U_{CE}=u^{-1}\left(\sum_x p_x u(x)\right).
\end{equation}
For a risk-averse agent, the certainty equivalent is smaller than the expected payoff; for a risk-seeking agent, it is larger.

In this paper the random payoff is the extracted work. Since work is additive over independent repetitions, the relevant risk-sensitive utility class is constant absolute risk aversion. We use the exponential utility
\begin{equation}
u_r(W)=\frac{1-\exp(-r\beta W)}{r},
\qquad r\neq 0,
\end{equation}
with the risk-neutral limit
\begin{equation}
u_0(W)=\beta W .
\end{equation}
Here $r$ is the risk-sensitivity parameter. For $r>0$, the utility is concave and the agent is risk averse; for $r=0$, the agent is risk neutral; and for $r<0$, the utility is convex and the agent is risk seeking. Positive affine changes of $u_r$ do not affect the agent's preferences, so the above normalization is only a convenient choice. Thus the parameter $r$ controls how strongly the agent penalizes or favors work fluctuations. In the main text, optimizing this quantity over thermodynamic protocols selects the physical work-extraction protocol itself, rather than merely assigning a value to a fixed work distribution.

\section{General protocol for adversarial work extraction}
\label{GeneralWorkExtraction}
In this appendix, we derive the work formula for a general adversarial engine.
Let $x$ be a microscopic state that is not necessarily binary.  Suppose that the initial distribution is given by $P (x)$ and the initial energy level is given by $E^B (x)$.  We relate the energy level to Bob's distribution by $Q^B(x) = e^{- E^B(x)}/Z^B$ with $Z^B = \sum_x e^{- E^B(x)}$ being the normalization factor (partition function).  Note that we set $k_B T$ to unity throughout this section. 

Let us first remember the optimal work extraction protocol at the level of ensemble average: (i) Alice quenches the energy level to $E (x)$ that is defined through $P (x) = e^{- E(x)}/Z$ with $Z = \sum_x e^{- E(x)}$; (ii) Alice moves the energy level from $E (x)$ to $E^B (x)$ quasi-statically and isothermally. It is well known  that Alice extracts the work $D(P \| Q^B)$ on average by this protocol \cite{takahirosbook,ParrondoHorowitzSagawa2015}.

Now we turn to the adversarial scenario. 
The protocol that Alice performs is: (i) Alice quenches the energy level to $E^A (x)$ that is defined through $Q^A (x) = e^{- E^A(x)}/Z^A$ with $Z^A = \sum_x e^{- E^A(x)}$ for her choice of $Q^A$; (i') Alice lets the system thermalize; (ii) Alice moves the energy level from $E^A (x)$ to $E^B (x)$ quasi-statically and isothermally. 
The extracted work in step (i) is given by $E^B (x) - E^A(x)$, while in step (ii) $\ln Z^B - \ln Z^A$ (the change in the equilibrium free energy). We thus obtain
\begin{equation}
w(x) = E^B (x) - E^A(x) +\ln Z^B - \ln Z^A = \ln \left( \frac{Q^A(x)}{Q^B(x)} \right),
\end{equation}
which reproduces \ref{generalsinglework}.

By taking the average with respect to $P$, Eq. \eqref{adversarialengine} is also reproduced.





\section{Maximization of the utility function}\label{utilitymaximization}

In this appendix, we compute the maximizer of the utility function for risk parameter $r$. 
Recall that a rational gambler will set their bet $Q^{A}$ as to maximize Eq.~\eqref{generalsinglework_u2}.
Note that the first term on the right-hand side is a constant, so we need only focus on the second term. For $r>0$ (risk-aversion), the second term is negative, whilst if $r<0$ (risk-seeking), the second term is positive. This means that for a risk-averse individual, we minimize  
\begin{equation}
   \min_{Q^{A}} \sum_{x} P(x) \bigg(  \frac{Q^{A}(x)}{Q^{B}(x)}\bigg)^{-r} , 
\end{equation}
which is equivalent to minimizing
\begin{equation}
     \min_{Q^{A}} \ln(  \sum_{x} P(x) \bigg( \frac{Q^{A}(x)}{Q^{B}(x)} \bigg)^{-r}  ).
\end{equation}
On the other hand, for a risk-seeking individual, we maximize
\begin{equation}
   \max_{Q^{A}} \sum_{x} P(x) \bigg(  \frac{Q^{A}(x)}{Q^{B}(x)}\bigg)^{-r}, 
\end{equation}
which is equivalent to maximizing
\begin{equation}
     \max_{Q^{A}} \ln(  \sum_{x} P(x) \bigg( \frac{Q^{A}(x)}{Q^{B}(x)} \bigg)^{-r}  ) .
\end{equation}

In order to find the minimizer (resp. maximizer), we will use a lemma proved in \cite{classicalswisspeople}, which states that 
\begin{equation}\label{classicalswisstheorem}
    \ln( \sum_{x}P(x)  \bigg( \frac{Q^{A}(x)}{Q^{B}(x)} \bigg)^{-r} ) = -r \big( D_{\frac{1}{1+r}}(P||Q^{B}) - D_{1+r}(G^{r}||Q^{A}) \big),
\end{equation}
where 
\begin{equation}\label{rationalstrategy}
    G^{r}= \frac{P(x)^{\frac{1}{1+r}} Q^{B}(x)^{\frac{r}{1+r}}}{ \sum_{x'}P(x')^{\frac{1}{1+r}} Q^{B}(x')^{\frac{r}{1+r}}} .
\end{equation}
The minimizer (resp. maximizer) is therefore given by Eq.~\eqref{rationalstrategy}.

\section{Calculation of the certainty equivalent}\label{certaintyequivalentCALC}

Recall that two equivalent ways of characterizing risk preferences were through the concavity/convexity of the utility function and the certainty equivalent. As shown in \cite{ducuaraandpaul}, the certainty equivalent has a clear information-theoretic interpretation in a thermodynamic context. We will also find it useful here to calculate the certainty equivalent of work for each value of the risk aversion parameter $r$ below. 

Since the utility of outcome $x$ is given by Eq.~\eqref{generalsinglework_u}, we  calculate
\begin{align}\label{beginning}
     W_{CE} &= u_{r}^{-1}\big( \mathbb E( u_{r}(w (x)) ) \big) \\
     &= u_{r}^{-1}\bigg( \mathbb E \bigg(  \frac{1}{r}( 1-  \bigg(  \frac{Q^{A}(x)}{Q^{B}(x)}\bigg)^{-r} \bigg) \bigg)  \\ 
      &= u_{r}^{-1}\bigg(  \frac{1}{r} \bigg( 1- \sum_{x} P(x) \bigg(  \frac{Q^{A}(x)}{Q^{B}(x)}\bigg)^{-r} \bigg) \bigg) \\ 
   &= -\frac{1}{r} \ln \bigg( \sum_{x}P(x) \bigg(  \frac{Q^{A}(x)}{Q^{B}(x)}\bigg)^{-r}  \bigg) ,
\end{align}
where in the last line we have used that the inverse function of $u_{r}(x)$ is $u_{r}^{-1}(x)=-\frac{1}{r} \ln(1-rx)$.
Again from the result of \cite{classicalswisspeople}, we obtain
\begin{equation}
     W_{CE} = D_{\frac{1}{1+r}}(P||Q^{B}) - D_{1+r}(G^{r}||Q^{A}) .
\end{equation}
 We finally conclude Eq.~\eqref{eq:WorkCE},
when the expression for $Q^{A,r}$ is substituted into the expression for the certainty equivalent. 

Since $D_{\alpha}<D_{\beta}$ for $\alpha<\beta$, one sees that for risk-averse (resp. risk-seeking) individuals, the certainty equivalent decreases (resp. increases)  with increasing (resp. decreasing) risk aversion, as expected. 

In addition, the work certainty equivalent is smaller than $D_{1}$ (the nonequilibrium free energy) for a risk averse player, and equal to $D_{1}$ for a risk neutral player. In particular, for an extremely risk averse player ($r \to \infty$), the certainty equivalent tends to $D_{0}(P||Q^{B})$, which becomes zero when both distributions have the same support. 


\section{Extracted work as a function of strategy}\label{extractedworkcalc}



Our goal of this Appendix is to express Eq.~\eqref{adversarialengine} for the optimal strategy $Q^{A}$ which from now on we write as $Q^{A,r}$, to emphasize dependence on the risk parameter of the CARA utility function. We substitute the explicit form of the optimal strategy into the second term.

Let $\alpha = \frac{1}{1+r}$ for notational clarity.
The optimal strategy for a given risk parameter $r$ is given by
\begin{equation}\label{eq:optimal_strategy}
Q^{A,r}(x) = \frac{P(x)^{\alpha} Q^B(x)^{1-\alpha}}{Z(\alpha )}, 
\end{equation}
where
\begin{align}
     Z(\alpha) = \sum_{x'} P(x')^{\alpha} Q^B(x')^{1-\alpha} \label{eq:optimal_strategy}.
\end{align}
 Since the first divergence is independent of the strategy  $Q^{A,r}$, we compute the second divergence term:
\begin{align}
D(P || Q^{A,r}) &= \sum_x P(x) \ln \frac{P(x)}{Q^{A,r}(x)} \\
&= \sum_x P(x) \left[ \ln P(x) - \ln \left( \frac{P(x)^\alpha Q^B(x)^{1-\alpha}}{Z(\alpha) } \right) \right] \\
&= \sum_x P(x) \left[ \ln P(x) - \alpha \ln P(x) - (1-\alpha) \ln Q^B(x) + \ln Z(\alpha)  \right] \\
&= (1 - \alpha) \sum_x P(x) \ln \frac{P(x)}{Q^B(x)} + \ln Z(\alpha)  \\
&= (1 - \alpha) D(P || Q^B) + \ln Z(\alpha) .\label{eq:intermediate_div}
\end{align}

One can see that the normalization constant $Z(\alpha )$ is related to the Rényi divergence by:
\begin{equation}
\ln Z(\alpha) = (\alpha - 1) D_\alpha(P || Q^B). \label{eq:logZ}
\end{equation}
Substituting \eqref{eq:logZ} back into \eqref{eq:intermediate_div}:
\begin{align}
D(P || Q^{A,r}) &= (1 - \alpha) D(P || Q^B) + (\alpha - 1) D_\alpha(P || Q^B) \\
&= (1 - \alpha) \left[ D(P || Q^B) - D_\alpha(P || Q^B) \right].
\end{align}
Finally, we substitute this result back into the original expression for expected work~\eqref{adversarialengine}:
\begin{align}
     D(P || Q^B) - D(P || Q^{A,r}) 
    &= D(P || Q^B) - (1 - \alpha) \left[ D(P || Q^B) - D_{\alpha}(P || Q^B) \right] \\
    &= D(P || Q^B) - (1 - \alpha) D(P || Q^B) + (1 - \alpha) D_\alpha(P || Q^B) \\
    &= \alpha D(P || Q^B) + (1 - \alpha) D_\alpha(P || Q^B).
\end{align}
Thus, the expected work extracted for the optimal strategy corresponding to risk parameter $r$ is given by Eq.~\eqref{eq:expected}.

\section{Conditions for rationality for $r<-1$}\label{theannoyingone}

In Eq.~\eqref{eq:WorkCE}, we showed that the certainty equivalent for work extraction for a rational agent with CARA utility function is a Rényi divergence.
The work extracted in a specific outcome $x$ is
\begin{equation}
w(x) =  k_{B}T\ln\left( \frac{Q^{A, r}(x)}{Q^B(x)} \right),
\end{equation}
where $Q^{A, r}$ is the optimal strategy for a given value of the risk parameter $r$.
Substituting this to the certainty equivalent, we get
\begin{equation}\label{violation_condition}
D_{\alpha}(P || Q^B) < \min_x \ln\left( \frac{Q^{A, r}(x)}{Q^B(x)} \right) .
\end{equation}
This is the general condition. We can now use the specific form of $Q^{A, r}$ to simplify the right-hand side. From  Eq.~\eqref{eq:optimal_strategy} and Eq.~\eqref{eq:logZ}, we obtain
\begin{equation}
w(x) = \ln\left( \frac{P(x)^{\alpha} Q^B(x)^{1-\alpha}}{Z(\alpha) Q^B(x)} \right) = \ln\left( \frac{P(x)^{\alpha} Q^B(x)^{-\alpha}}{Z(\alpha)} \right) = \alpha \ln\left( \frac{P(x)}{Q^B(x)} \right) - (\alpha - 1 ) D_{\alpha}(P || Q^B).
\end{equation}

To find the minimum value of this expression, recall that $\alpha < 0$ for $r<-1$.
Therefore,
\begin{equation}
\min_x w(x) = \alpha \cdot \max_x \ln\left( \frac{P(x)}{Q^B(x)} \right) - (\alpha - 1 ) D_{\alpha}(P || Q^B).
\end{equation}
We substitute this back into our violation condition \eqref{violation_condition} and obtain
\begin{equation}
D_{\alpha}(P || Q^B) < \alpha \cdot \max_x \ln\left( \frac{P(x)}{Q^B(x)} \right) - (\alpha - 1 ) D_{\alpha}(P || Q^B),
\end{equation}
or equivalently,
\begin{equation}
\alpha D_{\alpha}(P || Q^B) < \alpha \cdot \max_x \ln\left( \frac{P(x)}{Q^B(x)} \right).
\end{equation}
Again noticing that $\alpha < 0$, the violation condition \eqref{violation_condition} reduces to
\begin{equation}
D_{\alpha}(P || Q^B) > \max_x \ln\left( \frac{P(x)}{Q^B(x)} \right),
\end{equation}

Since \(D_{\alpha}(P\|Q^B)\leq D_{\infty}(P\|Q^B)\), this condition never holds. Thus rationality is never violated.

\section{Derivation of the work bound}
\label{appendix_work_bound}

In this Appendix, we show the derivation of Eq.~\eqref{work_bound_n}.
Recall that the optimal distribution $Q^{A*,\mu}(x)$ is given by Eq.~\eqref{optimalbet}, for which we define $Z(\mu ) = \sum_{x'} P(x')^\mu Q^B(x')^{1-\mu}$.
We can show that
\begin{equation}
    D(Q^{A*,\mu} \| Q^B) = \mu \sum_x Q^{A*,\mu} \ln \frac{P(x)}{Q^B(x)} -\ln Z(\mu),
\end{equation}
\begin{equation}
    D(Q^{A*,\mu} \| P) = - ( 1 - \mu ) \sum_x Q^{A*,\mu} \ln \frac{P(x)}{Q^B(x)} -\ln Z(\mu).
\end{equation}
By noting that $D_\mu (P \| Q^B) = - \frac{1}{1-\mu} \ln Z(\mu )$, we obtain
\begin{equation}
     D(Q^{A*,\mu} \| Q^B) = D_\mu (P \| Q^B) - \frac{\mu}{1-\mu} D(Q^{A*,\mu} \| P).
\end{equation}
Now recall that $Q^{A*,\mu} = \lambda_{x^n}$.
Because of the risk constraint \eqref{risk_constraint},
\begin{equation}
    D(Q^{A*,\mu} \| Q^B) \geq D_\mu (P \| Q^B) - \frac{\mu}{1-\mu} \frac{1}{n}\ln \frac{1}{\epsilon} .
\end{equation}
Going back to the work expression~\eqref{finite_n_main}, we obtain
\begin{equation}
    W_{n} (x^n) \geq  k_{B}T nD(Q^{A*,\mu}||Q^{B}) \geq k_{B}T n \left( D_\mu (P \| Q^B) - \frac{\mu}{1-\mu} \frac{1}{n}\ln \frac{1}{\epsilon} \right),
\end{equation}
which proves the bound.

\section{Reliability-constrained work extraction}
\label{appendix_work_bound}

Here we show that optimizing the reliability of a prescribed work-extraction
rate selects the same family of protocols as the CARA expected-utility
optimization in the main text. Throughout this section, we consider fixed protocols $Q^A$, with full support, applied identically in each of \(n\) independent rounds, as in the adversarial Szilard engine of the main text. We assume that the prior $P$ and reference distribution $Q^B$ are
distinct and have full support on a finite outcome set $\mathcal X$.
All logarithms in this section are natural.

For an outcome sequence $x^n=(x_1,\ldots,x_n)$, let $\lambda_{x^n}$ denote
its empirical distribution. The extracted work is
\begin{equation}
  W_n(x^n)=nk_BT\,G_{Q^A}(\lambda_{x^n}),
  \qquad
  G_{Q^A}(\lambda)
   :=\sum_x\lambda(x)\ln\frac{Q^A(x)}{Q^B(x)}
   =D(\lambda\|Q^B)-D(\lambda\|Q^A).
  \label{eq:sm_work_type}
\end{equation}
We fix a dimensionless work target $R>0$ and consider the probability of
extracting at least $nk_BT R$. Write
\begin{align}
  p_{\mathrm s}^{(n)}(R,Q^A)
    &:=P^{\otimes n}\!\left\{x^n:W_n(x^n)\ge nk_BT R\right\},
    \label{eq:sm_success_probability}\\
  p_{\mathrm f}^{(n)}(R,Q^A)
    &:=1-p_{\mathrm s}^{(n)}(R,Q^A).
    \label{eq:sm_failure_probability}
\end{align}
The maximum average work per round, in units of $k_BT$, is
$K:=D(P\|Q^B)$, attained by $Q^A=P$. For $0<R<K$, the relevant
reliability objective is to maximize the exponential decay rate of the
failure probability. For targets above $K$, success is a rare event and
we instead minimize its exponential decay rate.

For a fixed full-support protocol $Q^A$, the method of types gives the
corresponding probability exponents:
\begin{align}
  J_{\mathrm f}(R,Q^A)
  &:=\lim_{n\to\infty}-\frac1n\ln p_{\mathrm f}^{(n)}(R,Q^A)
   =\inf_{\lambda:G_{Q^A}(\lambda)\le R}D(\lambda\|P),
   \label{eq:sm_failure_exponent}\\
  J_{\mathrm s}(R,Q^A)
  &:=\lim_{n\to\infty}-\frac1n\ln p_{\mathrm s}^{(n)}(R,Q^A)
   =\inf_{\lambda:G_{Q^A}(\lambda)\ge R}D(\lambda\|P).
   \label{eq:sm_success_exponent}
\end{align}
The boundary conventions in these variational formulas follow by
approximating the relevant half-spaces with empirical types. In
particular, the success formula applies to the nondegenerate thresholds
considered below; empty event sets have exponent $+\infty$.

Define the tilted protocol, for $\mu>0$, by
\begin{equation}
  Q^{A*,\mu}(x)
   :=\frac{P(x)^\mu Q^B(x)^{1-\mu}}{Z_\mu},
  \qquad
  Z_\mu:=\sum_{y\in\mathcal X}P(y)^\mu Q^B(y)^{1-\mu},
  \label{eq:sm_tilted_protocol}
\end{equation}
and associate with it the target rate and probability exponent
\begin{equation}
  R_\mu:=D(Q^{A*,\mu}\|Q^B),\qquad
  d_\mu:=D(Q^{A*,\mu}\|P).
  \label{eq:sm_R_d}
\end{equation}
The parameter $\mu\in(0,1)$ covers $R_\mu\in(0,K)$, while $\mu>1$
covers $R_\mu\in(K,R_\infty)$, where
\begin{equation}
  R_\infty=-\ln Q^B(\mathcal M),\qquad
  \mathcal M:=\mathop{\mathrm{arg\,max}}_{x\in\mathcal X}
                 \frac{P(x)}{Q^B(x)}.
  \label{eq:sm_max_rate}
\end{equation}
Here $Q^B(\mathcal M)=\sum_{x\in\mathcal M}Q^B(x)$.
To see that these ranges are covered, let $\psi(\mu)=\ln Z_\mu$.
Then
\begin{equation}
  \frac{\mathrm dR_\mu}{\mathrm d\mu}
  =\mu\psi''(\mu)
  =\mu\,\mathrm{Var}_{Q^{A*,\mu}}
          \!\left[\ln\frac{P(X)}{Q^B(X)}\right]>0,
  \label{eq:sm_R_monotone}
\end{equation}
where strict positivity follows from $P\ne Q^B$ and full support.
The endpoints are $R_0=0$, $R_1=K$ and
$\lim_{\mu\to\infty}R_\mu=R_\infty$.

The central identity, valid for every probability distribution
$\lambda$ on $\mathcal X$ and every $\mu>0$, is
\begin{align}
 D(\lambda\|P)-d_\mu
 &=\frac1\mu D(\lambda\|Q^{A*,\mu})
    +\frac{\mu-1}{\mu}\bigl[D(\lambda\|Q^B)-R_\mu\bigr]
    \label{eq:sm_divergence_identity}\\
 &=D(\lambda\|Q^{A*,\mu})
    +\frac{\mu-1}{\mu}
       \bigl[G_{Q^{A*,\mu}}(\lambda)-R_\mu\bigr].
    \label{eq:sm_payoff_identity}
\end{align}
It follows directly by substituting the expression for
$Q^{A*,\mu}$ into the relative entropies and using
$G_{Q^A}(\lambda)=D(\lambda\|Q^B)-D(\lambda\|Q^A)$.

\paragraph{Targets below the average-work optimum.}
Let $0<\mu<1$ and fix $R=R_\mu<K$. For any protocol $Q^A$,
\begin{equation}
  G_{Q^A}(\lambda)\le D(\lambda\|Q^B),
\end{equation}
so its failure region contains the set
$\{\lambda:D(\lambda\|Q^B)\le R_\mu\}$.
Because $(\mu-1)/\mu<0$, Eq.~\eqref{eq:sm_divergence_identity}
shows that
\begin{equation}
  \inf_{\lambda:D(\lambda\|Q^B)\le R_\mu}
       D(\lambda\|P)=d_\mu,
\end{equation}
with equality at $\lambda=Q^{A*,\mu}$. Hence
\begin{equation}
  J_{\mathrm f}(R_\mu,Q^A)\le d_\mu
  \quad\text{for every fixed protocol }Q^A.
\end{equation}
For the tilted protocol itself, Eq.~\eqref{eq:sm_payoff_identity}
shows that $D(\lambda\|P)\ge d_\mu$ throughout its failure
region $G_{Q^{A*,\mu}}(\lambda)\le R_\mu$, with equality at
$\lambda=Q^{A*,\mu}$. Therefore
\begin{equation}
  \boxed{\sup_{Q^A}J_{\mathrm f}(R_\mu,Q^A)
         =J_{\mathrm f}(R_\mu,Q^{A*,\mu})=d_\mu,\qquad 0<\mu<1.}
  \label{eq:sm_below_K_optimum}
\end{equation}
Equivalently, this protocol has
$p_{\mathrm f}^{(n)}(R_\mu,Q^{A*,\mu})
 =\exp[-nd_\mu+o(n)]$.

\paragraph{Targets above the average-work optimum.}
Let $\mu>1$ and fix $R=R_\mu\in(K,R_\infty)$.
For any protocol $Q^A$, its success region is contained in
$\{\lambda:D(\lambda\|Q^B)\ge R_\mu\}$.
Now $(\mu-1)/\mu>0$, and Eq.~\eqref{eq:sm_divergence_identity}
shows that
\begin{equation}
  \inf_{\lambda:D(\lambda\|Q^B)\ge R_\mu}
       D(\lambda\|P)=d_\mu,
\end{equation}
again with equality at $\lambda=Q^{A*,\mu}$. Thus
$J_{\mathrm s}(R_\mu,Q^A)\ge d_\mu$ for every fixed protocol.
For $Q^A=Q^{A*,\mu}$, Eq.~\eqref{eq:sm_payoff_identity} gives the
reverse bound on the infimum, with equality at the same distribution.
Consequently,
\begin{equation}
  \boxed{\inf_{Q^A}J_{\mathrm s}(R_\mu,Q^A)
        =J_{\mathrm s}(R_\mu,Q^{A*,\mu})=d_\mu,\qquad \mu>1,}
  \label{eq:sm_above_K_optimum}
\end{equation}
and
$p_{\mathrm s}^{(n)}(R_\mu,Q^{A*,\mu})
 =\exp[-nd_\mu+o(n)]$.

\paragraph{Relation to risk-sensitive utility and R\'enyi divergences.}
The optimal protocols in Eqs.~\eqref{eq:sm_below_K_optimum}
and~\eqref{eq:sm_above_K_optimum} coincide with the CARA-optimal
protocols derived in the main text under the identification
\begin{equation}
  \boxed{\mu=\frac{1}{1+r}.}
  \label{eq:sm_r_mu}
\end{equation}
Thus $r>0$ selects the below-$K$ reliability branch and
$-1<r<0$ selects the above-$K$ branch. At $r=0$ ($\mu=1$),
$Q^{A*,1}=P$ recovers the average-work optimum. These results
concern the interior tilted family; they do not require an extension
to boundary-optimal protocols at stronger risk seeking.

Finally, since $\ln Z_\mu=(\mu-1)D_\mu(P\|Q^B)$, Eq.~\eqref{eq:sm_R_d}
implies the exact identity
\begin{equation}
  \boxed{R_\mu=D_\mu(P\|Q^B)
           +\frac{\mu}{\mu-1}\,d_\mu,\qquad \mu>0,\ \mu\ne1.}
  \label{eq:sm_renyi_frontier}
\end{equation}
For $0<\mu<1$, $d_\mu$ is the optimal failure exponent;
for $\mu>1$ in the stated target range, it is the optimal success
exponent. Equation~\eqref{eq:sm_renyi_frontier} therefore expresses
the reliability--yield frontier in terms of R\'enyi divergences.
The exponent statements hold at leading exponential order in $n$;
they do not assert exact finite-$n$ success probabilities.


\end{document}